\documentclass[12pt]{article}
\usepackage{graphicx}
\usepackage{subfigure}
\usepackage{placeins}
\usepackage{amsmath}
\usepackage{array}
\usepackage{multirow}
\usepackage{amssymb}
\usepackage{verbatim}
\usepackage{authblk}
\usepackage{slashed}

\newcommand{\tr}{{\rm tr}}

\newcommand{\gd}{\partial^a_{x,\mu}}
\newcommand{\gdr}{\tilde\partial^a_{x,\mu}}
\newcommand{\gdrprev}{\tilde\partial^a_{x-\hat\mu,\mu}}
\newcommand{\link}{U_{x,\mu}}

\newcommand{\staple}{L_{x,\mu}}

\newcommand{\MSbar}{\overline{\text{MS}}}
\newcommand{\refeq}[1]{Eq.~(\ref{#1})}

\numberwithin{equation}{section}

\title{Nonperturbative renormalization of the $\Delta S = 1$ weak Hamiltonian
including the $G_1$ operator}

\author{Greg McGlynn\thanks{gem2128@columbia.edu} }
\affil{Physics Department, Columbia University, New York, NY 10027, USA}

\date{\vspace{-5ex}}

\begin{document}
\maketitle

\begin{abstract}

Under renormalization, physical operators can mix with operators which vanish
by the equations of motion. Such operators cannot contribute to matrix elements
between physical states, but they contribute to operator mixing in
renormalization schemes which are defined at an off-shell momentum point, such
as the popular regularization-invariant schemes. For the first time, we
renormalize the lattice $\Delta S=1$ effective weak Hamiltonian taking into
account the most important such operator, $G_1 \propto \overline s \gamma_\nu
(1-\gamma_5) D_\mu G_{\mu\nu} d$.  This removes an important systematic error
in calculations of weak matrix elements on the lattice. 

\end{abstract}

\section{Introduction}

An important goal of lattice quantum chromodynamics (QCD) is the calculation of
weak transitions between hadronic states. The standard way to do this this is
to represent the weak interaction by an effective Hamiltonian and use the
lattice to calculate the matrix elements of this Hamiltonian. In the effective
Hamiltonian all particles much heavier than the lattice cutoff have been
integrated out.

The effective weak Hamiltonian is computed in continuum perturbation theory,
and the lattice Hamiltonian must be matched to the perturbative one. A
convenient set of matching conditions is given by the family of regularization
invariant (RI) renormalization schemes \cite{OriginalRI, RISMOM,
RISMOMPrecursor}. In these schemes we calculate the amputated Green's functions
of operators at a large off-shell Euclidean momentum point and require the
results to agree between the lattice and the continuum.

A curious aspect of this procedure is that operators which would usually vanish
by the equations of motion can mix with the physical operators. That is, in
off-shell Green's functions new divergences can appear which can only be
canceled by operators which vanish by the equations of motion. This happens
because the equations of motion are not valid in off-shell Green's functions
\cite{KlubergSternZuber, DeansDixon, JoglekarLee, Dawson1997}.

In this paper we focus on the $\Delta S = 1$ sector of the effective weak
Hamiltonian, which governs the important $K \to \pi\pi$ decay. At one loop, the
$\Delta S = 1$ weak Hamiltonian mixes with an operator $G_1$ which would
normally be redundant by the equations of motion. In perturbative calculations
of the renormalization of the weak Hamiltonian, the $G_1$ operator has been
taken into account \cite{ChristophMatchingFactors, BurasG123}. However, so far
$G_1$ has not been included on the lattice side of the matching calculation
\cite{OldRBCKPiPi, RBCUKQCDK2PiPi2015}. Previous calculations of lattice
renormalization factors for the effective weak Hamiltonian have therefore not
been strictly correct. 

In this paper we give two ways of including the $G_1$ operator in the
nonperturbative renormalization (NPR) of the lattice effective weak
Hamiltonian. Our methods have the advantage that the effect of the $G_1$
operator is confined to the NPR procedure. Once the NPR has been carried out,
we can forget about the $G_1$ operator, since it is after all redundant by the
equations of motion. Its effect is encoded in the renormalization factor matrix
of the physical operators.

\subsection{Operator basis}

We work in a three-flavor effective theory where all particles heavier than the
strange quark have been integrated out. We briefly review the operator basis
for the $\Delta S = 1$ three-flavor effective Hamiltonian, following
\cite{ChristophMatchingFactors}. This effective Hamiltonian can be written as a
linear combination of a basis of ten four-quark operators
\cite{EffectiveWeakHamiltonian}. All operators have the structure of the
product of two currents, have dimension 6, and have $\overline s d$ flavor
quantum numbers. The ten-operator basis is traditionally given as

\begin{equation}
\begin{split}
Q_1 &= (\bar s_i d_i)_{V-A} (\bar u_j u_j)_{V-A} \\
Q_2 &= (\bar s_i d_j)_{V-A} (\bar u_j u_i)_{V-A} \\
Q_3 &= (\bar s_i d_i)_{V-A} \sum_{q=u,d,s} (\bar q_j q_j)_{V-A} \\
Q_4 &= (\bar s_i d_j)_{V-A} \sum_{q=u,d,s} (\bar q_j q_i)_{V-A} \\
Q_5 &= (\bar s_i d_i)_{V-A} \sum_{q=u,d,s} (\bar q_j q_j)_{V+A} \\
Q_6 &= (\bar s_i d_j)_{V-A} \sum_{q=u,d,s} (\bar q_j q_i)_{V+A} \\
Q_7 &= \frac{3}{2} (\bar s_i d_i)_{V-A} \sum_{q=u,d,s} e_q (\bar q_j q_j)_{V+A} \\
Q_8 &= \frac{3}{2} (\bar s_i d_j)_{V-A} \sum_{q=u,d,s} e_q (\bar q_j q_i)_{V+A} \\
Q_9 &= \frac{3}{2} (\bar s_i d_i)_{V-A} \sum_{q=u,d,s} e_q (\bar q_j q_j)_{V-A} \\
Q_{10} &= \frac{3}{2} (\bar s_i d_j)_{V-A} \sum_{q=u,d,s} e_q (\bar q_j q_i)_{V-A}
\end{split}
\end{equation}

\noindent Here $i,j$ are color indices. The $V-A$ and $V+A$ subscripts denote
left and right-handed currents:

\begin{equation}
\begin{split}
(\bar q q)_{V-A} \equiv \overline q \gamma_\mu (1 - \gamma_5) q \\
(\bar q q)_{V+A} \equiv \overline q \gamma_\mu (1 + \gamma_5) q
\end{split}
\end{equation}

The effective Hamiltonian is then a linear combination

\begin{equation}
\mathcal H^\text{eff} = \sum_{i=1}^{10} w_i(\mu) Q_i(\mu)
\end{equation}

\noindent The $w_i(\mu)$ are Wilson coefficients. Both the Wilson coefficients
and the operators depend on the renormalization scheme and the renormalization
scale $\mu$. Continuum perturbation theory calculations usually renormalize the
four-quark operators in the $\MSbar$ scheme. 

Actually, this ten-operator basis is linearly dependent. The following
identities hold:

\begin{equation} \label{eq:TenOpBasisLinDep}
\begin{split}
Q_4 &= Q_2 + Q_3 - Q_1 \\
Q_9 &= \frac{3}{2}Q_1 - \frac{1}{2}Q_3 \\
Q_{10} &= \frac{1}{2}(Q_1 - Q_3) + Q_2
\end{split}
\end{equation}

\noindent Using these we can reduce our ten-operator basis to a seven-operator
basis, usually written

\begin{equation}
\begin{split}
Q'_1 &= 3Q_1 + 2Q_2 - Q_3 \\
Q'_2 &= \frac{1}{5}(2Q_1 - 2Q_2 + Q_3) \\
Q'_3 &= \frac{1}{5}(-3Q_1 + 3Q_2 + Q_3) \\
Q'_5 &= Q_5 \\
Q'_6 &= Q_6 \\
Q'_7 &= Q_7 \\
Q'_8 &= Q_8
\end{split}
\end{equation}

\noindent Note that there is no $Q'_4$. This seven-operator basis has the
advantage that each operator transforms in a definite way under the $SU(3)_L
\otimes SU(3)_R$ chiral symmetry group of massless QCD. The operators fall into
three representations of this group:

\begin{equation}
\begin{split}
Q'_1 &\in (27, 1) \\
Q'_{2,3,5,6} &\in (8, 1) \\
Q'_{7,8} &\in (8, 8)
\end{split}
\end{equation}

\noindent This seven-operator basis is called the chiral basis. When chiral
symmetry is preserved, for example when domain wall fermions are used, mixing
between operators in different representations is forbidden.

There is an eighth operator \cite{ChristophMatchingFactors}

\begin{equation} \label{eq:DefineG1}
\begin{split}
G_1 & \equiv \frac{4}{ig^2} \overline s \gamma_\mu (1 - \gamma_5) [D_\nu, [D_\nu, D_\mu]] d \\
& = -\frac{4}{g^2} \overline s \gamma_\mu (1 - \gamma_5) (D_\nu G_{\nu\mu}) d
\end{split}
\end{equation}

\noindent which is also dimension-6, has the same flavor quantum numbers as the
$\Delta S = 1$ operators, and transforms in the $(8, 1)$ representation of
$SU(3)_L \otimes SU(3)_R$. In the second line of \refeq{eq:DefineG1}, the
covariant derivative is in the adjoint representation: $D_\nu G_{\nu\mu} \equiv
\partial_\nu G_{\nu\mu} - i [A_\nu, G_{\nu\mu}]$.

In the RI schemes, this operator mixes with the four $(8, 1)$ four-quark
operators at one loop \cite{ChristophMatchingFactors}. There are even more
operators of this sort which mix at even higher loops; for example
\cite{BurasG123}

\begin{equation}
\begin{split}
G_2 & = \overline s \gamma_\mu (1 - \gamma_5) \{D^2, D_\mu\} d \\
G_3 & = \overline s D_\mu D_\nu D_\lambda (\gamma_\mu \gamma_\nu \gamma_\lambda - \gamma_\lambda \gamma_\nu \gamma_\mu) d
\end{split}
\end{equation}

\noindent We will only consider $G_1$, the only such operator to appear at one
loop.

In continuum QCD, the equation of motion of the gauge field is

\begin{equation} \label{eq:ContinuumEOM}
-\frac{1}{g^2} (D_\nu G_{\nu\mu}) = T^a \sum_q \overline q \gamma_\mu T^a q
\end{equation}

\noindent Here $a$ is an adjoint color index, $T^a$ are the su(3) generators,
and $q$ runs over the quark flavors. This can be used to rewrite $G_1$ as

\begin{equation} \label{eq:G1OnShellContinuum}
G_1 = Q_4 + Q_6 - \frac{1}{3}(Q_3 + Q_5) \equiv Q_p
\end{equation}

\noindent So it seems that $G_1$ is actually a linear combination of four-quark
operators in disguise. However, this is misleading. Inside a correlation
function, the equation of motion is only valid as an operator equation if there
are no contact terms. In the Green's functions used in the RI renormalization
schemes, contact terms spoil the equation of motion of the gauge field. One way
of understanding this is that the quark external states used in the RI schemes
are not gauge invariant. We deal with this by fixing Landau gauge, which can be
thought of as making the Green's function gauge invariant by contracting the
quark fields with Wilson lines. These Wilson lines can coincide with the
operator we are renormalizing, producing contact terms if we try to use the
equation of motion of the gauge field in that operator.

Therefore when we use the RI schemes we should treat $G_1$ as linearly
independent from the seven four-quark operators of the chiral basis, and we
should compute the mixing matrix of this expanded eight-operator basis.

While we need to include $G_1$ in our renormalization procedure, when we
compute physical matrix elements of the renormalized effective weak
Hamiltonian, we \emph{can} use the equations of motion. The reason is that in
physical matrix elements, the gauge-invariant operators that create and destroy
the external states are separated by some finite distance from the weak
Hamiltonian operator.  Therefore we do not have to worry about contact terms.
So in that step $G_1$ really is redundant and we ought to be able to eliminate
it using a lattice analog of \refeq{eq:G1OnShellContinuum}.

\subsection{Regularization invariant schemes}

The RI schemes for nonperturbative renormalization are a class of procedures
for constructing renormalized operators $O^{RI}_i(\mu)$ from bare lattice
operators $O^{lat}_i(a)$. We will work in the 8-operator basis $\{O_i\} =
\{Q'_1, \ldots, Q'_8, G_1\}$. The relation between the renormalized and bare
operators is given by the $8 \times 8$ Z-factor matrix:

\begin{equation} \label{eq:DefZ8x8}
O^{RI}_i(\mu) = Z^{lat \to RI}_{ij}(\mu, a) O^{lat}_j(a)
\end{equation}

We will work in the RI/SMOM scheme for the $\Delta S = 1$ operators, which is
defined in detail in \cite{ChristophMatchingFactors}. It is relatively
straightforward to implement this scheme on the lattice and obtain the $8
\times 8$  matrix $Z^{lat \to RI}$. 

We extend the RI/SMOM scheme defined in \cite{ChristophMatchingFactors} in two
small ways. First, we modify Eq.~(88) of \cite{ChristophMatchingFactors} so
that the projector $P_{4p,G_1}$ has both a even parity part and an odd parity
part; we use

\begin{equation}
P_{4p,G_1} = \delta_{ij}\delta_{kl}[(\gamma^\mu)(1-\gamma_5)]_{\beta\alpha}(\gamma_\mu)_{\delta\gamma}
\end{equation}

\noindent This allows us to perform the NPR procedure using only the
parity-even parts of the operators or using only the parity-odd parts of the
operators.

Second, while \cite{ChristophMatchingFactors} only defines renormalization
conditions for the four-quark operators $Q'_i$, we also construct a
renormalized $G_1$. The renormalization conditions for $G_1$ are exactly
analogous to those for the four-quark operators: the projected amputated
Green's functions of the renormalized operator $G_1^{RI}$ in the eight external
states are chosen to be equal to their tree level values.

Once we have constructed the renormalized operators, we compute weak transition
amplitudes for the physical process $i \to f$ as a linear combination of the
matrix elements

\begin{equation}
\langle f | O^{RI}_i(\mu) | i \rangle = Z_{ij}^{lat \to RI}(\mu, a) \langle f | O^{lat}_j(a) | i \rangle
\end{equation}

\noindent In such matrix elements, the equations of motion are valid and so we
ought to be able to simplify these expressions by eliminating $G_1$ using the
equations of motion. In particular, it should never be necessary to calculate
$\langle f | G_1^{lat} | i\rangle$. In the rest of this paper we give two
methods of achieving this simplification.

\section{Method 1: Eliminating $G_1$ with perturbation theory}
\label{sec:PerturbativeMethod}

To distinguish the seven four-quark operators from $G_1$, we can split
\refeq{eq:DefZ8x8} into two equations:

\begin{equation} \label{eq:QprimeRI}
Q'^{RI}_i(\mu) = Z^{lat \to RI, 7 \times 7}_{ij}(\mu, a) Q'^{lat}_j(a) + c_i^{lat \to RI}(\mu, a) G_1^{lat}(a)
\end{equation}
\begin{equation} \label{eq:G1RI}
G_1^{RI}(\mu) = d^{lat \to RI}_i(\mu, a) Q'^{lat}_i(a) + Z_{G_1}^{lat \to RI}(\mu, a) G_1^{lat}(a)
\end{equation}

\noindent Here the $Q'_i$ are the seven four-quark operators, $Z^{lat \to RI, 7
\times 7}$ is the $7 \times 7$ block of the Z-factor matrix that gives the
mixing among the four-quark operators, and $c_i$, $d_i$, and $Z_{G_1}$ are the
rest of the $8 \times 8$ Z-factor matrix, which deal with $G_1$.

In the absence of contact terms, $G_1^{RI}(\mu)$ can be replaced by a
linear combination of the four-quark operators $Q'^{RI}_i(\mu)$ in matrix
elements:

\begin{equation} \label{eq:IntroduceSFactors}
\langle f | G_1^{RI}(\mu) | i \rangle = s_i(\mu) \langle f | Q'^{RI}_i(\mu) | i \rangle
\end{equation}

The coefficients $s_i(\mu)$ can be expanded in a perturbation series. To tree
level they are

\begin{equation} \label{eq:DefineSFactors}
\left( \begin{array}{c}
s_1 \\ s_2 \\ s_3 \\ s_5 \\ s_6 \\ s_7 \\ s_8
\end{array} \right)
=
\left( \begin{array}{c}
0 \\ 1 \\ 7/3 \\ -1/3 \\ 1 \\ 0 \\ 0
\end{array} \right)
+ O(\alpha_s(\mu))
\end{equation}

\noindent These tree level values can be found from \refeq{eq:TenOpBasisLinDep}
and \refeq{eq:G1OnShellContinuum}. Those equations give us the tree level
relation because they are written in terms of bare continuum operators. 

Now using \refeq{eq:QprimeRI} and \refeq{eq:G1RI} in
\refeq{eq:IntroduceSFactors} we obtain

\begin{equation} 
\langle G_1^{lat}(a) \rangle = k_j(\mu,a) \langle Q'^{lat}_j(a) \rangle
\end{equation}

\begin{equation} \label{eq:DefineKFactors}
k_j(\mu,a) \equiv 
\frac{s_i(\mu) Z_{ij}^{lat \to RI}(\mu,a) - d_j^{lat \to RI}(\mu,a)}
{Z_{G_1}^{lat \to RI}(\mu,a) - s_k(\mu) c_k^{lat \to RI}(\mu,a)}
\end{equation}

\noindent This equation tells us how to eliminate the lattice operator
$G_1^{lat}(a)$ in favor of the four-quark lattice operators $Q'^{lat}_i(a)$
when we have to compute a physical matrix element. It is a lattice analog of
\refeq{eq:G1OnShellContinuum}.

Now suppose we want to compute a physical
matrix element of a renormalized RI four-quark operator. Using
\refeq{eq:DefineKFactors} in \refeq{eq:QprimeRI} gives

\begin{equation} \label{eq:DeriveRMatrix}
\langle Q'^{RI}_i(\mu) \rangle = R^{lat \to RI}_{ij}(\mu,a) \langle Q'^{lat}_j(a) \rangle
\end{equation}

\begin{equation} \label{eq:DefineRMatrix}
R^{lat \to RI}_{ij}(\mu, a) \equiv Z^{lat \to RI}_{ij}(\mu,a) + c_i^{lat \to RI}(\mu,a) k_j(\mu, a)
\end{equation}

\noindent This $R$ matrix tells us how to compute a physical matrix element of
a renormalized four-quark operator solely in terms of lattice four-quark
operators, without having to compute physical matrix elements of $G_1^{lat}$.
Of course, we will still have to compute momentum-space Green's functions of
$G_1^{lat}$ to carry out the NPR procedure. But after that, the effect of $G_1$
on the renormalized four-quark operators is captured by the $c_i$ and $k_i$
factors. Once we have these and construct the $R^{lat \to RI}$ matrix, we can
forget about $G_1$.

This strategy is very convenient because in our application it is currently
only necessary to compute the coefficients $s_i$ at tree level. That is, we do
not have to calculate the $O(\alpha_s)$ corrections to
\refeq{eq:DefineSFactors}. The reason is that before we can use the $R^{lat \to
RI}$ matrix, we have to do another conversion from the RI renormalization
scheme to the $\MSbar$ renormalization scheme, because the effective weak
Hamiltonian is constructed in the $\MSbar$ scheme in continuum perturbation
theory. We end up with a lattice to $\MSbar$ conversion matrix:

\begin{equation}
R^{lat \to \MSbar} = R^{RI \to \MSbar} \times R^{lat \to RI}
\end{equation}

\noindent The $R^{RI \to \MSbar}$ matrix has been computed perturbatively in
\cite{ChristophMatchingFactors}, but only to one loop. This limits our
calculation to one-loop accuracy, so we can consistently neglect two loop
effects. In \refeq{eq:DefineRMatrix}, the $s_i$ appear inside $k_j$, which
multiplies $c_i$. The quantity $c_i$ starts at one loop, so the one loop
correction to $s_i$ is a two-loop effect and can be neglected in our
calculation. 

We will check after the fact that the change due to including $G_1$ in the NPR
is numerically fairly small. One-loop corrections to $s_i$ would produce a
small change in this small change, so their overall effect is very small and
can be neglected.

For us to trust the preceding argument, we should be using
\refeq{eq:DefineSFactors} at a reasonably high energy scale, such that we
believe the one-loop corrections to \refeq{eq:DefineSFactors} are indeed
substantially smaller than the tree level values. 

\subsection{$G_1$ lattice operator}

On the lattice there are many possible discretizations of any continuum
operator. In the calculations presented below we use 

\begin{equation}
G_1^{lat}(x) = \bar s_x \gamma_\mu (1 - \gamma_5) B_{x,\mu} d_x
\end{equation}

\noindent Here $B_{x,\mu}$ is a discretization of $D_\nu G_{\nu\mu}(x)$.
Inspired by the continuum equation of motion \refeq{eq:ContinuumEOM}, we choose
this discretization to be related to the lattice gauge field equation of
motion. Given a lattice gauge action $S_g^{lat}(U)$, we define

\begin{equation} \label{eq:DefineB}
B_{x,\mu} \equiv \frac{3}{\beta} i T^a (\gd + \gdrprev) S_g^{lat}(U) \\
\end{equation}

\noindent where the link derivatives are defined by

\begin{equation}
\gd f(\link) \equiv \frac{d}{ds} f(e^{i s T^a} \link)|_{s = 0}
\end{equation}

\begin{equation}
\gdr f(\link) \equiv \frac{d}{ds} f(\link e^{i s T^a})|_{s = 0}
\end{equation}

\noindent and our convention for the generators $T^a$ is $\tr[T^a T^b] =
\delta^{ab}/2$. The combination of $\gd$ and $\gdrprev$ ensures that
$B_{x,\mu}$ transforms in a definite way under parity. \refeq{eq:DefineB} is
the lattice analog of the continuum equation

\begin{equation}
D_\nu G_{\nu\mu}(x) \propto \frac{\delta}{\delta A_\mu(x)} S_g^\text{cont}(A)
\end{equation}

\noindent where $S_g^\text{cont}$ is the pure Yang-Mills continuum action.

Explicitly, our $G_1$ operator is

\begin{equation}
G_1^{lat}(x) = \frac{1}{2} \overline s_x \gamma_\mu (1 - \gamma_5) [\link \staple + L_{x-\hat\mu,\mu} U_{x-\hat\mu,\mu}]_{TA} d_x
\end{equation}

\noindent where $L_{x,\mu}$ is commonly called the ``staple'' for the gauge
action $S_g(U)$. The notation $[ \cdot ]_{TA}$ denotes the traceless
antihermitian part of a matrix. In the calculations below, $S_g(U)$ is the
Iwasaki gauge action.

\subsection{Results}

As an example of this strategy, we compute the $Z^{lat \to RI}$ and $R^{lat \to
RI}$ matrices on a $24^3 \times 64$ 2+1 flavor Shamir domain wall ensemble with
$m_l = 0.005$, $m_s = 0.04$, $L_s = 16$ \cite{RBC2432}. The gauge action is the
Iwasaki action with $\beta = 2.13$. The lattice spacing is $a^{-1} =
1.7848(50)$ GeV \cite{LatestEnsemblePaper}. We use the momenta

\begin{equation}
\begin{split}
a p_1 & = \frac{2\pi}{24}(2, 4, -2, 0) \\
a p_2 & = \frac{2\pi}{24}(4, 2, 2, 0) \\
\mu & = |p_1| = |p_2| = 2.29 \text{ GeV}
\end{split}
\end{equation}

\noindent We use a valence mass $a m_l = a m_s = 0.01$. We measure on 792
configurations. We use only the parity-odd parts of the operators and
projectors. We do not measure the wave function renormalization $Z_q$, so we
only give results up to a factor of $Z_q^{-2}$. We work in the
RI/SMOM($\gamma_\mu$, $\gamma_\mu$) scheme of \cite{ChristophMatchingFactors}.
We find

\footnotesize
\begin{equation}
\begin{split}
\hspace{-1.2in}&\hspace{1.2in}Z_q^{-2} Z^{lat \to RI} = \\
\hspace{-1.2in}&\left(\begin{array}{rrrrrrrr}
0.846179(42) & & & & & & & \\
& 0.9400(37) & -0.0860(20) & -0.0025(17) & 0.00076(81) & & & -0.0090(42) \\
& -0.0850(12) & 0.94007(93) & -0.00155(60) & -0.00076(29) & & & 0.0509(23) \\
& -0.028(12) & -0.0193(62) & 0.9659(48) & -0.1422(23) & & & -0.005(13) \\
& -0.0037(39) & 0.0034(28) & -0.0532(17) & 0.70199(92) & & & 0.1470(70) \\
& & & & & 0.959102(32) & -0.142791(16) & \\
& & & & & -0.052603(11) & 0.703316(50) & \\
& -0.007(31) & 0.248(22) & -0.147(16) & -0.0259(74) & & & 2.301(59) \\
\end{array}\right)
\end{split}
\end{equation}
\normalsize

\noindent The upper-left $1 \times 1$ block corresponds to the $(27,1)$
operator $Q'_1$.  The next $4 \times 4$ block corresponds to the $(8, 1)$
operators $Q'_{2,3,5,6}$.  The next $2 \times 2$ block corresponds to the $(8,
8)$ operators $Q'_{7,8}$. The last $1 \times 1$ block corresponds to $G_1$.
Entries equal to zero have been omitted. In this and what follows, all quoted
errors are statistical only.

We see that there is no mixing between different representations of $SU(3)_L
\otimes SU(3)_R$ and that $G_1$ mixes with the $(8,1)$ operators, as expected.
We can read off the $c_i$'s of \refeq{eq:QprimeRI} from the last column:

\begin{equation}
\begin{split}
Z_q^{-2} c_2 & = -0.0090(42) \\
Z_q^{-2} c_3 & = 0.0509(23) \\
Z_q^{-2} c_5 & = -0.005(13) \\
Z_q^{-2} c_6 & = 0.1470(70) \\
\end{split}
\end{equation}

\noindent These tell us how much $G_1^{lat}$ appears in each RI four-quark
operator. We see that $G_1$ mainly mixes with $Q'_3$ and $Q'_6$, and the
corresponding $c_i$'s are measured to about 5\%.

Having found the $8 \times 8$ Z-factor matrix, we next eliminate $G_1$
by computing the $k_i$'s of \refeq{eq:DefineKFactors}. Using
the tree-level values of $s_i$ from \refeq{eq:DefineSFactors}, we find

\begin{equation}
\begin{split}
k_2 & = 0.370(18) \\
k_3 & = 0.915(28) \\
k_5 & = -0.115(8) \\
k_6 & = 0.379(10) \\
\end{split}
\end{equation}

Finally we get the $7 \times 7$ $R^{lat \to RI}$ matrix:

\footnotesize
\begin{equation}
\begin{split}
\hspace{-0.6in}&\hspace{0.6in}Z_q^{-2} R^{lat \to RI} = \\
\hspace{-0.6in}&\left(\begin{array}{rrrrrrr}
0.846179(42) & & & & & & \\
& 0.9367(45) & -0.0942(51) & -0.0015(16) & -0.0026(15) & & \\
& -0.0661(16) & 0.9867(24) & -0.00738(66) & 0.01855(83) & & \\
& -0.029(14) & -0.024(15) & 0.9665(46) & -0.1440(45) & & \\
& 0.0506(53) & 0.1379(76) & -0.0700(18) & 0.7577(24) & & \\
& & & & & 0.959102(32) & -0.142791(16) \\
& & & & & -0.052603(11) & 0.703316(50) \\
\end{array} \right)
\end{split}
\end{equation}
\normalsize

We would like to understand what effect the inclusion of $G_1$ has had on our
final answer. So we also carry out the NPR procedure using only the original
7-operator basis, neglecting $G_1$, and compute the difference between our
$R^{lat \to RI}$ matrix and the $7 \times 7$ Z-factor matrix obtained by
neglecting $G_1$:

\begin{equation} \label{eq:DefineDeltaR}
\Delta R^{lat \to RI} \equiv R^{lat \to RI} - Z^{lat \to RI, 7 \times 7, \text{ no } G_1}
\end{equation}

\noindent We find

\begin{equation} \label{eq:24cubedG1Difference}
\begin{split}
&Z_q^{-2} \Delta R^{lat \to RI} = \\
&\left(\begin{array}{rrrrrrr}
0 & & & & & & \\
& -0.00090(41) & -0.0025(12) & 0.00044(20) & -0.00090(42) & & \\
& 0.00510(24) & 0.01423(65) & -0.00248(11) & 0.00512(23) & & \\
& -0.0005(13) & -0.0013(36) & 0.00023(63) & -0.0005(13) & & \\
& 0.01470(76) & 0.0410(20) & -0.00717(35) & 0.01476(73) & & \\
& & & & & 0 & 0 \\
& & & & & 0 & 0 \\
\end{array} \right)
\end{split}
\end{equation}

\noindent Of course, only the $(8, 1)$ sub-block of the $R^{lat \to RI}$ matrix
is affected. We find that the biggest effect is in the third and fifth rows
corresponding to $Q^{'RI}_3$ and $Q^{'RI}_6$. The effect of including $G_1$ is
clearly resolved.

The fact that $\Delta R$, the change due to $G_1$, is fairly small compared to
the overall matrix $R$ reassures us that we are justified in neglecting
one-loop corrections to the $s_i$ coefficients. Those one-loop corrections
would only produce a small change in the already small matrix $\Delta R$.

\subsection{Step scaling for $K \to \pi\pi$} \label{sec:K2PiPiStepScaling}

The principal motivation for nonperturbative renormalization of the $\Delta S =
1$ weak Hamiltonian on the lattice is the calculation of the $K \to \pi\pi$
decay. In the recent RBC/UKQCD calculation in \cite{RBCUKQCDK2PiPi2015}, the
largest single systematic error came from operator renormalization.  We now
renormalize the $\Delta S = 1$ Hamiltonian for this calculation, including the
effects of $G_1$.

The $K \to \pi\pi$ calculation was carried out on a relatively coarse lattice
with $a^{-1} \approx 1.38$ GeV. The ensemble parameters are those of the DSDR
lattices described in \cite{DSDRPaper} except that in the $K \to \pi\pi$
calculation the quark masses are physical. We carry out the NPR procedure on
the $a m_l = 0.001$ ensemble of \cite{DSDRPaper}.

Because this lattice is quite coarse, we cannot perform the NPR at a very high
energy scale. This is a problem because the perturbative step we rely on to
convert from the RI/SMOM scheme to the $\MSbar$ scheme is not reliable at this
low energy. To solve this, we perform a nonperturbative step scaling
calculation on the $24^3$ ensemble used above.

We briefly describe the step scaling procedure. Suppose we have a lattice with
a relatively coarse lattice spacing $a_\text{coarse}$. If we do nonperturbative
renormalization on this lattice we can only construct RI operators at some
relatively low energy scale $\mu_\text{low}$. We use an auxiliary lattice with
a finer lattice spacing $a_\text{fine} < a_\text{coarse}$ on which we can do
NPR both at the scale $\mu_\text{low}$ and at a higher scale $\mu_\text{high}$.
We use this lattice to find a relation between the RI operators at the low
scale and the high scale, as follows:

\begin{equation}
\begin{split}
O^{RI}_i(\mu_\text{low}) &= Z^{lat \to RI}_{ij}(\mu_\text{low}, a_\text{fine}) O^{lat}_j(a_\text{fine}) \\
O^{RI}_i(\mu_\text{high}) &= Z^{lat \to RI}_{ij}(\mu_\text{high}, a_\text{fine}) O^{lat}_j(a_\text{fine}) \\
O^{RI}_i(\mu_\text{high}) &= \Sigma_{ij}(\mu_\text{high}, \mu_\text{low}) O^{RI}_j(\mu_\text{low})
\end{split}
\end{equation}

\noindent where the step scaling matrix is calculated as 

\begin{equation} \label{eq:DefineSigma}
\Sigma_{ij}(\mu_\text{high}, \mu_\text{low}) 
= Z^{lat \to RI}_{ik}(\mu_\text{high}, a_\text{fine}) [Z^{lat \to RI}(\mu_\text{low}, a_\text{fine})]^{-1}_{kj}
\end{equation}

\noindent Finally we can construct operators renormalized at the high scale in
terms of lattice operators defined on the coarse lattice:

\begin{equation}
O^{RI}_i(\mu_\text{high}) = Z^{lat \to RI}_{ij}(\mu_\text{high}, a_\text{coarse}) O^{lat}_j(a_\text{coarse})
\end{equation}

\noindent where

\begin{equation}
Z^{lat \to RI}_{ij}(\mu_\text{high}, a_\text{coarse}) 
\equiv \Sigma_{ij}(\mu_\text{high}, \mu_\text{low}) Z^{lat \to RI}_{jk}(\mu_\text{low}, a_\text{coarse})
\end{equation}

When we include $G_1$ in the NPR, the step-scaling should be carried out with
the $8 \times 8$ Z-factor matrices, giving an $8 \times 8$ step-scaling matrix
$\Sigma$ and eventually an $8 \times 8$ Z-factor matrix $Z^{lat \to
RI}(\mu_\text{high}, a_\text{coarse})$. From this Z-factor matrix, we can
construct the on-shell conversion matrix $R^{lat \to RI}(\mu_\text{high},
a_\text{coarse})$ using \refeq{eq:DefineRMatrix}.

We use the $24^3$ ensemble considered above as the fine lattice for step
scaling. So we have $a_\text{coarse}^{-1} \approx 1.37$ GeV, the lattice
spacing of the $32^3$ DSDR ensemble, and $a_\text{fine}^{-1} \approx 1.78$ GeV,
the lattice spacing of the finer $24^3$ Iwasaki ensemble. We use
$\mu_\text{low} = 1.33$ GeV and $\mu_\text{high} = 2.29$ GeV. 

On the coarse $32^3$ lattice we average over two sets of momenta for the external states:

\begin{equation}
\begin{split}
a_\text{coarse} p_{1,\text{low},a} & = \frac{2\pi}{32}(-4, -2, 2, 0) \\
a_\text{coarse} p_{2,\text{low},a} & = \frac{2\pi}{32}(0, -2, 4, -2) \\
a_\text{coarse} p_{1,\text{low},b} & = \frac{2\pi}{32}(0, 4, -2, -2) \\
a_\text{coarse} p_{2,\text{low},b} & = \frac{2\pi}{32}(-2, 0, -2, -4) \\
\end{split}
\end{equation}

\noindent These satisfy 

\begin{equation} 
\begin{split}
&p_{1,\text{low},a}^2 = p_{2,\text{low},a}^2 = (p_{1,\text{low},a} - p_{2,\text{low},a})^2 \\
&= p_{1,\text{low},b}^2 = p_{2,\text{low},b}^2 = (p_{1,\text{low},b} - p_{2,\text{low},b})^2 = \mu_\text{low}^2 
\end{split}
\end{equation}

\noindent (The calculation on the coarse lattice dominates the statistical
error, so we average over two sets of momenta to increase our statistics). On
the fine $24^3$ lattice we use the momenta

\begin{equation}
\begin{split}
a_\text{fine} p_{1,\text{low}} & = \frac{2\pi}{24}(0, 2, 2, 0) \\
a_\text{fine} p_{2,\text{low}} & = \frac{2\pi}{24}(2, 2, 0, 0) \\
a_\text{fine} p_{1,\text{high}} & = \frac{2\pi}{24}(2, 4, -2, 0) \\
a_\text{fine} p_{2,\text{high}} & = \frac{2\pi}{24}(4, 2, 2, 0) \\
\end{split}
\end{equation}

We use 792 configurations on the $24^3$ ensemble and 350 configurations on the $32^3$ ensemble.

We perform the NPR for two renormalization schemes, called
RI/SMOM($\gamma_\mu$, $\gamma_\mu$) and RI/SMOM($\slashed q$, $\slashed q$),
both defined in \cite{ChristophMatchingFactors}. These schemes are based on
different sets of projected Green's functions and use different wave function
renormalizations. 

For each scheme we give the step-scaled $8 \times 8$ Z-factor matrix and the
resulting $R^{lat \to RI}$ matrix to be used on-shell.  Again we can ask what
the effect of $G_1$ has been on the calculation. To do this we redo the entire
step scaling calculation with $7 \times 7$ Z-factor matrices neglecting $G_1$
(not shown below) and compute the $\Delta R$ matrix defined in
\refeq{eq:DefineDeltaR}. The errors we quote are purely statistical and do not
include the uncertainty in the wave function renormalization.

In the RI/SMOM($\gamma_\mu$, $\gamma_\mu$) scheme we find

\footnotesize
\begin{equation}
\begin{split}
\hspace{-1.2in}&\hspace{1.2in}
Z^{lat \to RI}(\mu_\text{high},a_\text{coarse}) = \\
\hspace{-1.2in}&\left(\begin{array}{rrrrrrrr}
   0.404278(54) &                &                &                &                &                &                &                \\
                &     0.4753(79) &    -0.0584(76) &    -0.0038(33) &    -0.0001(16) &                &                &      0.028(24) \\
                &    -0.0722(44) &     0.4620(45) &    -0.0020(18) &    0.00210(96) &                &                &      0.017(15) \\
                &      0.005(22) &      0.011(22) &     0.4593(91) &    -0.0598(49) &                &                &      0.064(66) \\
                &      0.005(16) &     -0.009(16) &    -0.0432(68) &     0.4008(34) &                &                &      0.091(53) \\
                &                &                &                &                &   0.470405(61) &  -0.062295(40) &                \\
                &                &                &                &                &  -0.034206(53) &    0.39486(15) &                \\
                &      -0.08(12) &      -0.02(12) &     -0.063(52) &      0.009(25) &                &                &       1.25(40) \\
\end{array}\right)
\end{split}
\end{equation}
\normalsize

\footnotesize
\begin{equation}
\begin{split}
\hspace{-1.2in}&\hspace{1.2in}
R^{lat \to RI}(\mu_\text{high},a_\text{coarse}) = \\
\hspace{-1.2in}&\left(\begin{array}{rrrrrrr}
   0.404278(54) &                &                &                &                &                &                \\
                &      0.485(13) &     -0.033(27) &    -0.0073(52) &     0.0104(84) &                &                \\
                &    -0.0663(82) &      0.478(17) &    -0.0041(29) &     0.0084(52) &                &                \\
                &      0.027(39) &      0.071(79) &      0.451(14) &     -0.036(23) &                &                \\
                &      0.036(29) &      0.074(62) &     -0.055(11) &      0.435(19) &                &                \\
                &                &                &                &                &   0.470405(61) &  -0.062295(40) \\
                &                &                &                &                &  -0.034206(53) &    0.39486(15) \\
\end{array}\right)
\end{split}
\end{equation}
\normalsize

\begin{equation} 
\begin{split}
&\Delta R^{lat \to RI} = \\
&\left(\begin{array}{rrrrrrr}
              0 &                &                &                &                &                &                \\
                &     0.0100(79) &      0.028(21) &    -0.0049(39) &     0.0089(70) &                &                \\
                &    -0.0012(47) &     -0.003(13) &     0.0003(24) &    -0.0002(43) &                &                \\
                &      0.021(22) &      0.058(59) &     -0.010(11) &      0.019(20) &                &                \\
                &      0.010(17) &      0.029(47) &    -0.0056(84) &      0.011(15) &                &                \\
                &                &                &                &                &              0 &              0 \\
                &                &                &                &                &              0 &              0 \\
\end{array} \right)
\end{split}
\end{equation}

In the RI/SMOM($\slashed q$, $\slashed q$) scheme we find

\footnotesize
\begin{equation}
\begin{split}
\hspace{-1.2in}&\hspace{1.2in}
Z^{lat \to RI}(\mu_\text{high},a_\text{coarse}) = \\
\hspace{-1.2in}&\left(\begin{array}{rrrrrrrr}
    0.42650(11) &                &                &                &                &                &                &                \\
                &      0.414(14) &     -0.188(16) &    -0.0005(56) &    -0.0025(31) &                &                &      0.017(22) \\
                &      0.043(15) &      0.698(18) &    -0.0053(62) &    -0.0000(39) &                &                &      0.032(28) \\
                &      0.029(47) &      0.051(53) &      0.464(18) &     -0.068(11) &                &                &      0.064(74) \\
                &      0.027(33) &      0.033(43) &     -0.056(17) &     0.4590(95) &                &                &      0.105(66) \\
                &                &                &                &                &    0.47130(15) &  -0.063099(59) &                \\
                &                &                &                &                &   -0.06136(11) &    0.45973(19) &                \\
                &       0.14(25) &       0.24(33) &      -0.19(12) &      0.069(73) &                &                &       1.45(50) \\
\end{array}\right)
\end{split}
\end{equation}
\normalsize

\footnotesize
\begin{equation}
\begin{split}
\hspace{-1.2in}&\hspace{1.2in}
R^{lat \to RI}(\mu_\text{high},a_\text{coarse}) = \\
\hspace{-1.2in}&\left(\begin{array}{rrrrrrr}
    0.42650(11) &                &                &                &                &                &                \\
                &      0.419(17) &     -0.171(31) &    -0.0010(59) &     0.0031(90) &                &                \\
                &      0.052(19) &      0.728(37) &    -0.0062(66) &      0.010(11) &                &                \\
                &      0.049(57) &       0.11(10) &      0.463(19) &     -0.048(31) &                &                \\
                &      0.059(41) &      0.133(91) &     -0.059(17) &      0.493(27) &                &                \\
                &                &                &                &                &    0.47130(15) &  -0.063099(59) \\
                &                &                &                &                &   -0.06136(11) &    0.45973(19) \\
\end{array}\right)
\end{split}
\end{equation}
\normalsize

\begin{equation} 
\begin{split}
&\Delta R^{lat \to RI} = \\
&\left(\begin{array}{rrrrrrr}
              0 &                &                &                &                &                &                \\
                &     0.0027(63) &      0.017(23) &    -0.0035(42) &     0.0061(79) &                &                \\
                &     0.0142(76) &      0.037(29) &    -0.0048(50) &     0.0120(97) &                &                \\
                &      0.016(21) &      0.066(79) &     -0.012(14) &      0.023(27) &                &                \\
                &      0.041(18) &      0.117(70) &     -0.017(12) &      0.039(24) &                &                \\
                &                &                &                &                &              0 &              0 \\
                &                &                &                &                &              0 &              0 \\
\end{array} \right)
\end{split}
\end{equation}

These $R$ matrices can be used to renormalize the operators used in the $K \to
\pi\pi$ calculation. Because the statistical errors are much larger on the
$32^3$ ensemble, the effect of $G_1$ on the calculation is not as easy to
resolve as it was in \refeq{eq:24cubedG1Difference}. The values given above for
$\Delta R$ bound the effect of $G_1$ on this calculation to be fairly small.

\section{Method 2: Use the lattice equation of motion}

In this section we describe an alternative strategy for using the RI scheme
with the $G_1$ operator. The strategy of Section \ref{sec:PerturbativeMethod}
relies on perturbation theory. We might be interested in a strategy that
produces the $R^{lat \to RI}$ matrix without any reliance on perturbation
theory. Here we give a strategy for doing this which uses the exact equation of
motion for the lattice gauge field.

The idea is that the eighth operator in our basis need not be the $G_1$
operator, but rather we can choose it to be a related operator which vanishes
exactly by the equation of motion. In the continuum, this related operator
would be $G_1 - Q_p$. On the lattice, the equation of motion of the gauge field
is

\begin{equation} \label{eq:LatticeEOM}
M_1^{lat} \equiv i T^a \gd S^{lat}(U, \psi) = 0
\end{equation}

\noindent where $S^{lat}$ is the total lattice action, $U$ is the gauge field, and $\psi$
represents the fermion fields. 

The operator $M_1^{lat}$ is the lattice analog of the operator $G_1 - Q_p$.
$M_1^{lat}$ vanishes exactly when the equations of motion are valid.
$M_1^{lat}$ does \emph{not} vanish in the momentum-space Green's functions used
to define the RI schemes. $M_1^{lat}$ can be used instead of $G_1^{lat}$ as the
eighth operator in our NPR basis.

This approach has advantages and disadvantages. The advantages are:

\begin{itemize}

\item Once we find $Z^{lat \to RI}$ there is no extra step needed to construct
$R^{lat \to RI}$, and in particular no need to eliminate $G_1^{lat}$ by a
perturbative calculation.

\item There is no need to define a renormalized $M_1^{RI}$ as we had to define
a renormalized $G_1^{RI}$. We only need the lattice operator $M_1^{lat}$.

\end{itemize}

\noindent To see how these advantages come about, write the seven RI four-quark
operators as

\begin{equation}
Q'^{RI}_i(\mu) = Z^{lat \to RI, 7 \times 7}(\mu, a) Q'^{lat}_i(a) + c^{lat \to RI}_i (\mu,a) M_1^{lat}(a)
\end{equation}

\noindent Here $i,j$ range only over the seven values $\{1, 2, 3, 5, 6, 7,
8\}$. To determine $Z^{lat \to RI, 7 \times 7}$ and $c^{lat \to RI}$ we impose
exactly the same conditions we imposed on the $Q'^{RI}_i(\mu)$ in Section
\ref{sec:PerturbativeMethod}. Suppose we have done this. Then when we compute a
physical matrix element of $Q'^{RI}_i(\mu)$, the $M_1^{lat}$ operator drops out
by the equation of motion:

\begin{equation}
\begin{split}
\langle f | Q'^{RI}_i(\mu) | i \rangle & = 
Z^{lat \to RI, 7 \times 7}(\mu, a) \langle f | Q'^{lat}_i(a) | i \rangle + 
c^{lat \to RI}_i(\mu, a) \langle f | M_1^{lat}(a) | i \rangle  \\
\langle f | Q'^{RI}_i(\mu) | i \rangle & = 
Z^{lat \to RI, 7 \times 7}(\mu, a) \langle f | Q'^{lat}_i(a) | i \rangle
\end{split}
\end{equation}

So in this strategy $Z^{lat \to RI, 7 \times 7}$ is the same matrix as $R^{lat
\to RI}$.  $M_1^{lat}$ drops out of physical matrix elements automatically, so
there is no need to do a perturbative calculation to eliminate it, so there is
no need to construct a renormalized $M_1^{RI}$ operator.

This is the same renormalization scheme as the one used in Section
\ref{sec:PerturbativeMethod}, because we impose the same renormalization
conditions on the four-quark operators $Q'^{RI}_i(\mu)$. We have merely chosen
a slightly different basis of lattice operators, which is our privilege. The
final results for $R^{lat \to RI}$ should differ only by lattice artifacts, and
because of the use of perturbation theory in finding the $s_i$'s in Section
\ref{sec:PerturbativeMethod}.

The disadvantage of this strategy is that the $M_1^{lat}$ operator is extremely
complicated.  Its exact form depends on the details of the full action used to
generate the ensemble, and it is usually quite messy.  Because of the link
derivative, $M_1^{lat}$ is point-split instead of local to one particular site.
In the case of domain wall fermions $M_1^{lat}$ involves five-dimensional
fermion fields, not just the fields on the four-dimensional boundary. This
makes the implementation of the contractions far more complicated and also
makes the computation much slower, since it is necessary to sum over the fifth
dimension.  Some ensembles we work with use a dislocation-suppressing
determinant ratio \cite{DSDRPaper}, which adds a new, complicated term to
$M_1^{lat}$. For these reasons we prefer the strategy of Section
\ref{sec:PerturbativeMethod} in practice. The strategy described in this
section is an option if the perturbative approximation of the $s_i$
coefficients is not acceptable.

\section{Summary}

It is an interesting aspect of quantum field theory that that when we
renormalize an operator it may mix with operators which vanish by the equations
of motion. This occurs already at one loop when we renormalize the $\Delta S =
1$ weak Hamiltonian. In this case, the $G_1$ operator appears and we need to
properly account for its mixing with the physical four-quark current-current
operators. Previous lattice calculations of the important $K \to \pi\pi$ decay
have neglected this effect. We have given two practical methods for including
the effects of $G_1$. 

The first method uses perturbation theory to compute the relation which holds
in physical matrix elements between the renormalized $G_1$ operator and the
renormalized four-quark operators. The second method uses instead the exact
lattice equation of motion. Both methods allow us to confine the effect of
$G_1$ to the NPR calculation, meaning that we never have to compute the lattice
matrix elements of $G_1$ between physical external states.

\section{Acknowledgements}

I thank Robert D. Mawhinney, Ziyuan Bai, and the rest of the RBC/UKQCD
collaboration for useful discussions. The calculations described here were
performed on Blue Gene/Q supercomputers at Brookhaven National Laboratory and
the Argonne Leadership Computing Facility. This work was supported in part by
U.S. DOE grant \#DE-SC0011941.


\begin{thebibliography}{99}

\bibitem{OriginalRI}
  G.~Martinelli, C.~Pittori, C.~T.~Sachrajda, M.~Testa and A.~Vladikas,
  ``A General method for nonperturbative renormalization of lattice operators,''
  Nucl.\ Phys.\ B {\bf 445}, 81 (1995)
  [hep-lat/9411010].

\bibitem{RISMOM} 
  C.~Sturm, Y.~Aoki, N.~H.~Christ, T.~Izubuchi, C.~T.~C.~Sachrajda and A.~Soni,
  ``Renormalization of quark bilinear operators in a momentum-subtraction scheme with a nonexceptional subtraction point,''
  Phys.\ Rev.\ D {\bf 80}, 014501 (2009)
  [arXiv:0901.2599 [hep-ph]].

\bibitem{RISMOMPrecursor} 
  Y.~Aoki {\it et al.},
  ``Non-perturbative renormalization of quark bilinear operators and B(K) using domain wall fermions,''
  Phys.\ Rev.\ D {\bf 78}, 054510 (2008)
  [arXiv:0712.1061 [hep-lat]].

\bibitem{KlubergSternZuber} 
  H.~Kluberg-Stern and J.~B.~Zuber,
  ``Renormalization of Nonabelian Gauge Theories in a Background Field Gauge. 2. Gauge Invariant Operators,''
  Phys.\ Rev.\ D {\bf 12}, 3159 (1975).
\bibitem{DeansDixon} 
  W.~S.~Deans and J.~A.~Dixon,
  ``Theory of Gauge Invariant Operators: Their Renormalization and S Matrix Elements,''
  Phys.\ Rev.\ D {\bf 18}, 1113 (1978).
\bibitem{JoglekarLee} 
  S.~D.~Joglekar and B.~W.~Lee,
  ``General Theory of Renormalization of Gauge Invariant Operators,''
  Annals Phys.\  {\bf 97}, 160 (1976).

\bibitem{Dawson1997} 
  C.~Dawson, G.~Martinelli, G.~C.~Rossi, C.~T.~Sachrajda, S.~R.~Sharpe, M.~Talevi and M.~Testa,
  ``New lattice approaches to the delta I = 1/2 rule,''
  Nucl.\ Phys.\ B {\bf 514}, 313 (1998)
  [hep-lat/9707009].

\bibitem{ChristophMatchingFactors} 
  C.~Lehner and C.~Sturm,
  ``Matching factors for Delta S=1 four-quark operators in RI/SMOM schemes,''
  Phys.\ Rev.\ D {\bf 84}, 014001 (2011)
  [arXiv:1104.4948 [hep-ph]].


\bibitem{BurasG123} 
  A.~J.~Buras, M.~Jamin, M.~E.~Lautenbacher and P.~H.~Weisz,
  ``Two loop anomalous dimension matrix for Delta S = 1 weak nonleptonic decays. 1. O(alpha-s**2),''
  Nucl.\ Phys.\ B {\bf 400}, 37 (1993)
  [hep-ph/9211304].

\bibitem{OldRBCKPiPi} 
  T.~Blum {\it et al.} [RBC Collaboration],
  ``Kaon matrix elements and CP violation from quenched lattice QCD: 1. The three flavor case,''
  Phys.\ Rev.\ D {\bf 68}, 114506 (2003)
  [hep-lat/0110075].

\bibitem{RBCUKQCDK2PiPi2015}
  Z.~Bai {\it et al.} [RBC and UKQCD Collaborations],
  ``Standard Model Prediction for Direct CP Violation in $K \to \pi\pi$ Decay,''
  Phys.\ Rev.\ Lett.\  {\bf 115} (2015) no.21,  212001
  [arXiv:1505.07863 [hep-lat]]. 
  Erratum: Z.~Bai {\it et al.},
  ``Erratum: Standard-model prediction for direct CP violation in $K\to\pi\pi$ decay,''
  arXiv:1603.03065 [hep-lat].

\bibitem{EffectiveWeakHamiltonian} 
  F.~J.~Gilman and M.~B.~Wise,
  ``Effective Hamiltonian for Delta s = 1 Weak Nonleptonic Decays in the Six Quark Model,''
  Phys.\ Rev.\ D {\bf 20}, 2392 (1979).


\bibitem{RBC2432} 
  Y.~Aoki {\it et al.} [RBC and UKQCD Collaborations],
  ``Continuum Limit Physics from 2+1 Flavor Domain Wall QCD,''
  Phys.\ Rev.\ D {\bf 83}, 074508 (2011)
  [arXiv:1011.0892 [hep-lat]].

\bibitem{LatestEnsemblePaper}
  T.~Blum {\it et al.} [RBC and UKQCD Collaborations],
  ``Domain wall QCD with physical quark masses,''
  Phys.\ Rev.\ D {\bf 93}, no. 7, 074505 (2016)
  [arXiv:1411.7017 [hep-lat]].

\bibitem{DSDRPaper}
  R.~Arthur {\it et al.} [RBC and UKQCD Collaborations],
  ``Domain Wall QCD with Near-Physical Pions,''
  Phys.\ Rev.\ D {\bf 87}, 094514 (2013)
  [arXiv:1208.4412 [hep-lat]].

\end{thebibliography}
\end{document}